\documentclass[runningheads]{llncs}
\usepackage{graphicx}
\usepackage{tabularray}
\usepackage{tikz}
\usepackage{amsmath}
\usepackage{breqn}
\newcommand{\LArrow}[1]{%
\parbox{#1}{\tikz{\draw[->](0,0)--(#1,0);}}
}

\begin{document}
\title{SCPAT-GAN: Structural Constrained and Pathology Aware Convolutional Transformer-GAN for Virtual Histology Staining of Human Coronary OCT images}
%
%
\author{Xueshen Li$^{1}$, Hongshan Liu$^{1}$, Xiaoyu Song$^{3}$, Brigitta C. Brott$^{2}$, Silvio H. Litovsky$^{2}$, and Yu Gan$^{1}$}
\authorrunning{X.Li et al.}
%
\institute{$^{1}$Department of Biomedical Engineering, Stevens Institute of Technology\\
$^{2}$School of Medicine, The University of Alabama at Birmingham\\
$^{3}$The Icahn School of Medicine at Mount Sinai\\
\email{\{ygan5\}@stevens.edu}}
\maketitle              
\begin{abstract}
There is a significant need for the generation of virtual histological information from coronary optical coherence tomography (OCT) images to better guide the treatment of coronary artery disease. However, existing methods either require a large pixel-wisely paired training dataset or have limited capability to map pathological regions. To address these issues, we proposed a structural constrained, pathology aware, transformer generative adversarial network, namely SCPAT-GAN, to generate virtual stained H\&E histology from OCT images. 
The proposed SCPAT-GAN advances existing methods via a novel design to impose pathological guidance on structural layers using transformer-based network. 
Our experiments on human coronary data have demonstrated the superiority of SCPAT-GAN in comparison with existing methods. 
Also, qualitative comparison and a blind test from a pathologist indicate that our virtual histology images are indistinguishable from real histology. 

\end{abstract}

\section{Introduction}
Coronary artery disease (CAD) is narrowing of coronary arteries caused by build-up of atherosclerotic plaques. As the most common type of heart disease, CAD leads to 1 in 7 deaths in the United States \cite{hajar2017risk}.
Optical Coherence Tomography (OCT) has been recognized as a valuable tool for imaging coronary tissue structures due to its high resolution capabilities \cite{Tearney.2012}. 
However, real-time interpretation of OCT images requires a significant amount of expertise and prior training. Additionally, the power of OCT interpretation, especially of the pathological region, is hindered by the lack of histopathological correlation. 
At present, direct histopathological analysis requires invasive and time-consuming evaluation that involves post-mortem tissue examination. 
The use of multiple reagents in histopathology can also lead to detrimental effects on tissue imaging. 
Histopathological analysis is not suitable for clinical use in patients, who require real-time tissue characterization of coronary arteries.

\begin{figure}[t]
\centering
\includegraphics[width=11cm]{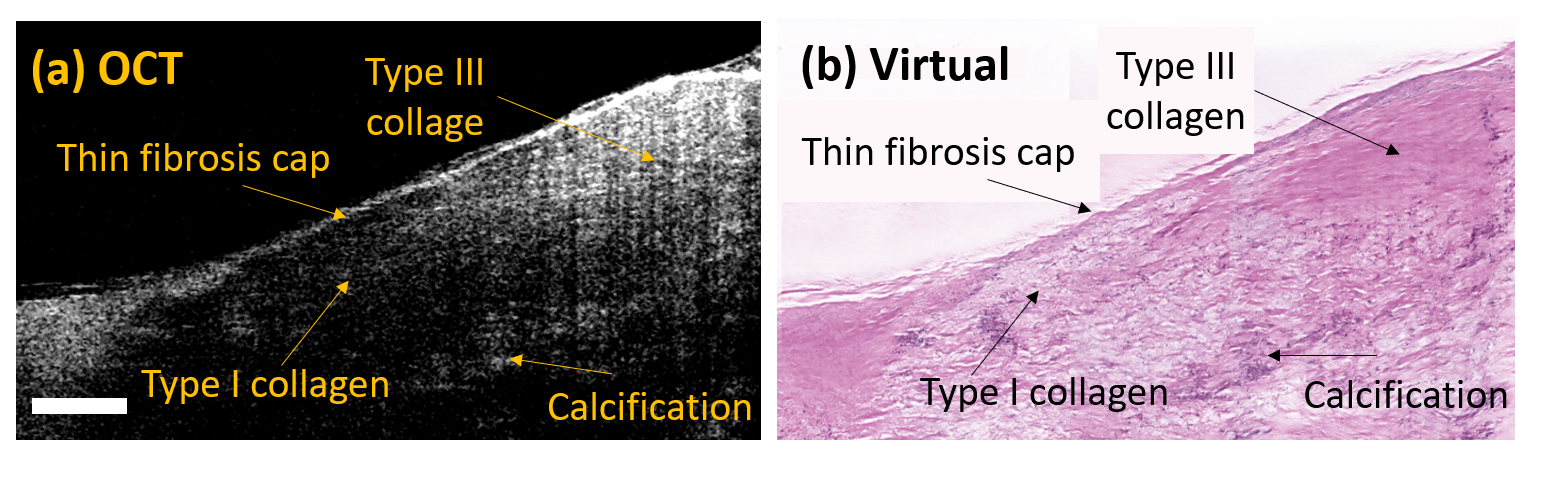}
\caption{(a) Example OCT image of a human coronary sample. (b) Virtual histology image generated from the example OCT images in (a). The scale bar: 500 $\mu$m. }
\label{fig:VirtualHistology}
\end{figure}

Incorporating histopathological visualization into real-time OCT imaging holds great potential to complement OCT with histopathological visualization. A typical example of generating virtual histology images from OCT images of human coronary artery is shown in Figure \ref{fig:VirtualHistology}. To date, there are limited frameworks developed to generate virtual histology from OCT images \cite{Winetraub.2021,https://doi.org/10.48550/arxiv.2211.06737}. Winetraub et al. used Pix2Pix Generative Adversarial Networks (GANs) to generate virtual stained Hematoxylin and Eosin (H\&E) histology for human skin tissues \cite{Winetraub.2021}. However, Pix2Pix GAN for virtual staining requires a pixel-wisely paired OCT and H\&E image dataset. The creation of a pixel-wisely paired dataset demands a significant investment of resources and labor, including the embedding of samples in fluorescent gel, photo-bleaching, and manual fine alignment \cite{Winetraub.2021}. Such a method also lacks generalizability to blood vessel which is a deformable soft tissue. Our previous method \cite{https://doi.org/10.48550/arxiv.2211.06737} demonstrates the capability to segment the three-layer structure (i.e., intima, media, and adventitia) in both OCT and H\&E images, thereby generating virtual H\&E images optimized for different layers in human coronary. However, current performance has not been optimal if there are pathological patterns, such as calcium and lipid accumulation, that alter the typical three-layer structure of human coronary arteries. 

To generate pathological-related regions from an unpaired dataset, we propose a \textbf{S}tructural \textbf{C}onstrained, \textbf{P}athology \textbf{A}ware, convolutional \textbf{T}ransformer GAN (SCPAT-GAN) to generate virtually stained H\&E histology images from OCT images. The proposed SCPAT-GAN incorporates two key components to enhance image quality for both normal and pathological coronary samples: a structural constraining module and a pathology awareness module. In summary, our main contributions include: 

(1) We propose a convolutional transformer-GAN structure for virtual H\&E staining of human coronary arteries based on OCT. This generative method does not require pixel-wisely mapping in the training dataset.

(2) We incorporate a structural constraining and pathology awareness modules for virtually staining coronary arteries with both normal three-layer structures and pathological patterns. 

(3) 
We conduct extensive experiments, including a blind test, to demonstrate that the generated images are indistinguishable from real histology images.
\section{Methodology}
\subsection{Design of SCPAT-GAN}

\subsubsection{Network architecture: }
\begin{figure}[t]
\centering
\includegraphics[width=12cm]{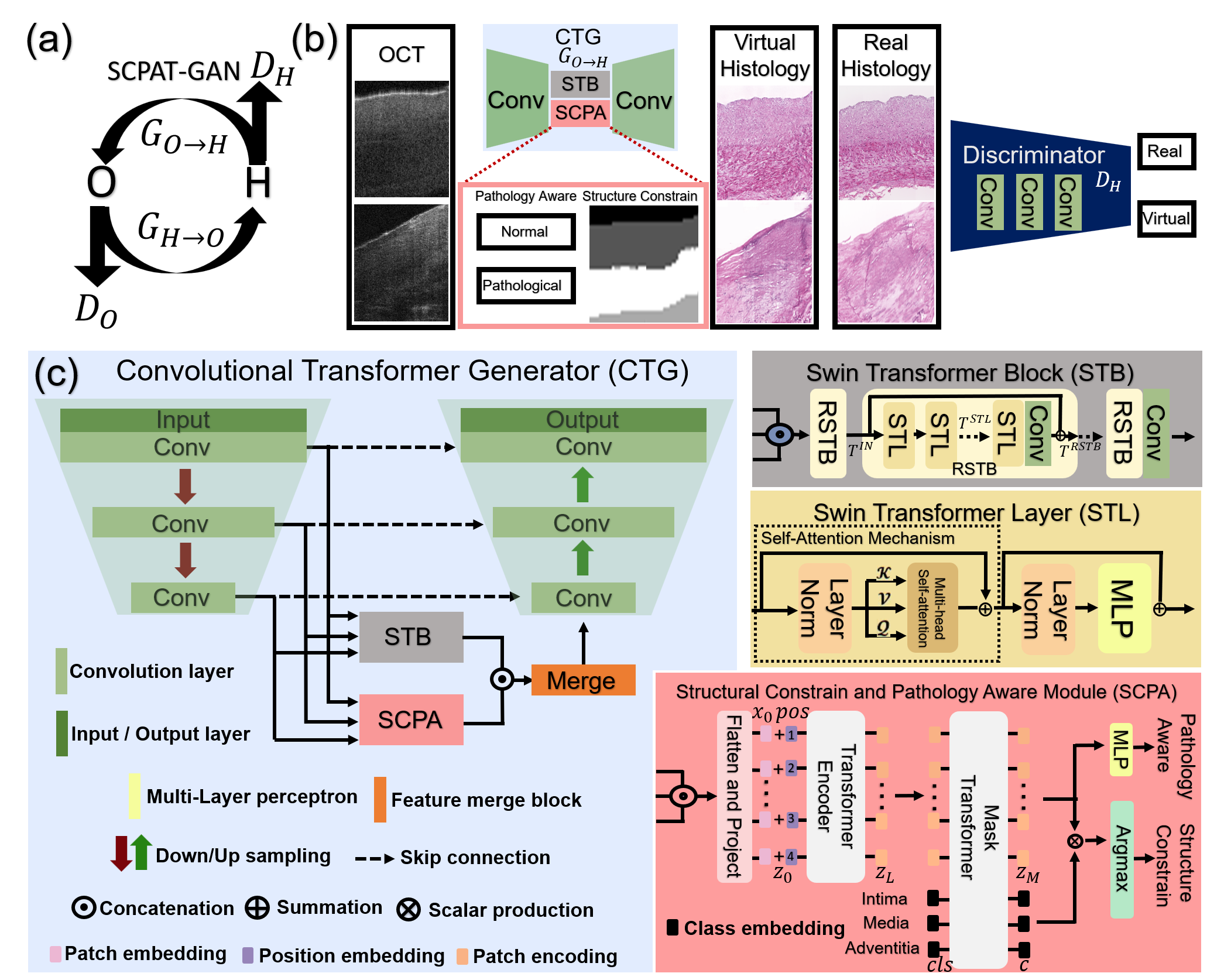}
\caption{(a): The design of the SCPAT-GAN. (b): The scheme of $G_{OH}$ and $D_H$. The $G_{OH}$ performs virtual staining based on OCT images. The $D_H$ distinguish the virtual histology images from real histology images. The SCPA module guide the virtual staining process by performing structural constraining and pathology awareness functions. (c): Details of the convolutional transformer generator (CTG). The multi-scale features are fed to STB and SCPA modules for virtual staining. }
\label{fig:SctrualSCPAT}
\end{figure}

The design of SCPAT-GAN is shown in Figure \ref{fig:SctrualSCPAT}. The SCPAT-GAN consists of two convolutional transformer generators ($G_{O\LArrow{.15cm}H}$ and G$_{H\LArrow{.15cm}O}$) and two discriminators ($D_H$ and $D_O$). \textcolor{black}{The transformer structure possesses self-focus mechanisms which provide the global context of a given data sample even at the lowest layer.} $G_{O\LArrow{.15cm}H}$ transfers images from OCT domain to histology domain; $G_{H\LArrow{.15cm}O}$ transfers images from the histology domain to OCT domain. The two generators share a similar structure. D$_H$ is the discriminator for histology images and D$_O$ is the discriminator for OCT images. Symbols $O$ and $H$ stand for OCT and histology images respectively.

The convolutional transformer generators ($G_{O\LArrow{.15cm}H}$ and G$_{H\LArrow{.15cm}O}$) take advantage of U-Net \cite{RFB15a} like structure to extract multi-scale features. The multi-scale features are sent to Swin Transformer Block (STB) and Structural Constraint and Pathology Aware (SCPA) Module.
The STB is a deep neural network architecture that employs multiple residual Swin Transformer sub-blocks (RSTBs) to extract features from input data. The RSTBs contain various Swin Transformer Layers (STLs) \cite{9607618} that facilitate local attention and cross-window interaction learning. 
The feature extraction process of RSTBs is expressed as: $T^{RSTB} = Conv(F^{STL} + T^{IN})$, where $F^{STL}$ denotes the features generated from STLs, Conv represents 2D convolutional layer with a kernel size of 3 × 3, and $T^{IN}$ represents the input feature of RSTBs. 
Each STL comprises components including layer normalization, multi-head self-attention (MLA) modules, residual connections, and a two-level multilayer perceptron (MLP) with GELU non-linearity. The self-attention of each head can be calculated as: $Attention(Q, K, V) = SoftMax(\frac{QK^{T}}{\sqrt{d}} + B)V$, where Q, K, V $\in {R}^{N^{2 \times d}}$ are the query, key and value matrices; $d$ denotes the query dimension; $N$ stands for the number of patches in a window; and $B \in {R}^{(2N-1)\times(2N+1)}$. 

\subsubsection{Structural constraining and pathology awareness: }
The SCPA module is based on a transformer encoder-decoder architecture, which guides the virtual staining procedure. The SCPA module performs structural constraining and pathology awareness functions by segmenting the human coronary layers and classifying the types of coronary samples (normal or pathological). 
The multi-scale features are split into a sequence of patches x=$[x_1, ..., x_N]\in R^{N\times P^2\times C}$, where $(P,P)$ stands for the patch size, and C is the number of channels of the multi-scale features. The patches are flattened and then linearly projected to an embedding sequence x$_0=[E_{x_1}, ..., E_{x_N}] \in R^{N\times d}$, where $d$ is the embedding dimension. Learnable position embeddings pos=$[pos_1, ..., pos_N] \in R^{N\times d}$ are added to the sequence of patch embeddings to generate the tokens z$_0$=x$_0$+pos for Encoder. The Encoder maps the input sequence z$_0$ to z$_L=[z_{L_1},..., z_{L_N}]$, which is an encoding sequence containing contextualized information of multi-scale features. 

The SCPA module is designed to be aware of pathology patterns as well as maintain and constrain the normal structure of coronary samples. In the case of normal coronary samples, the z$_L$ is decoded to a segmentation map \textbf{s} $\in R^{H \times W \times K}$, where $K=3$ and represents the three-layer structure of human coronary arteries. The segmentation map is acquired by the SCPA module, taking the scalar production between patch embeddings z$_M$ and class embeddings $c$: $Segmentaion = z_M c^T$, where z$_M$ is acquired by decoding z$_L$, and $c$ is acquired by decoding a set of three randomly initialized learnable class embeddings [cls$_{Intima}$, cls$_{Media}$, cls$_{Adventitia}$] corresponding to the three coronary layers. In the case of diseased coronary samples, the patch embeddings z$_M$ are sent to a two-level MLP for classification between normal and pathological coronary images: $Classification = MLP(z_M)$.
Also, the patch embeddings z$_M$ is concatenated to the extracted features from STB and then merged and up-sampled for OCT$\LArrow{.2cm}$Histology and Histology$\LArrow{.2cm}$OCT conversion. 

\subsubsection{Loss function: }
The loss function $L$ of SCPAT-GAN consists of five terms, which are adversarial loss $L_{adv}$, cycle-consistency loss $L_{cycle}$, embedding loss $L_{embedding}$, structural constraint loss $L_{SC}$, pathology awareness loss $L_{PA}$. 
\begin{equation} \label{eq1}
\begin{split}
&L(G_{O\LArrow{.15cm}H}, G_{H\LArrow{.15cm}O}, D_H, D_O, G^{SC}_{O\LArrow{.15cm}H},  G^{SC}_{H\LArrow{.15cm}O}, G^{PA}_{O\LArrow{.15cm}H}, G^{PA}_{H\LArrow{.15cm}O}) 
\\
& = L_{adv}(G_{O\LArrow{.15cm}H}, D_H) +  L_{adv}(G_{H\LArrow{.15cm}O}, D_O)
\\
&+ \alpha L_{cycle}(G_{O\LArrow{.15cm}H}, G_{H\LArrow{.15cm}O}) + \beta L_{embedding}(G_{O\LArrow{.15cm}H}, G_{H\LArrow{.15cm}O}) \\
&+ \gamma L_{SC}(G^{SC}_{O\LArrow{.15cm}H}, G^{SC}_{H\LArrow{.15cm}O}) + \iota L_{PA}(G^{PA}_{O\LArrow{.15cm}H}, G^{PA}_{H\LArrow{.15cm}O})
\end{split}
\end{equation}
We follow the definition of $L_{adv}$ and $L_{cycle}$ in \cite{8237506} and definition of $L_{embedding}$ in \cite{Liu.2021}. $\alpha$, $\beta$, $\gamma$, and $\iota$ are hyper-parameters. G$_{O\LArrow{.15cm}H}$ and G$_{H\LArrow{.15cm}O}$ are two generators that generate virtual histology images from OCT images and virtual OCT images from histology images respectively. G$^{SC}_{O\LArrow{.15cm}H}$,  G$^{SC}_{H\LArrow{.15cm}O}$, G$^{PA}_{O\LArrow{.15cm}H}$,  and G$^{PA}_{H\LArrow{.15cm}O}$ are the SCPA modules for performing structural constraining and pathology awareness functions in the generators. 
The $L_{SC}$ is implemented by segmentation loss:
\begin{equation} \label{eq2}
\begin{split}
L_{SC} = E_H[-S_H^{-1}\sum^{s_H}_{n=1}\sum^{C}_{c=1}y^{n,c}_H\log(G^{ SC}_{O\LArrow{.15cm}H}(O))] 
 +E_O[-S_O^{-1}\sum^{s_O}_{n=1}\sum^{C}_{c=1}y^{n,c}_O\log(G^{SC}_{H\LArrow{.15cm}O}(H))]
\end{split}
\end{equation}
$S_H$ and $S_O$ stand for the number of pixels in segmentation maps. $y^{n,c}_H$ and $y^{n,c}_O$ are the ground-truth pixel labels of different coronary layers for H\&E and OCT images respectively. $C$ stands for the number of categories of the coronary layers (c=3). The $L_{PA}$ is implemented by classification loss:
\begin{equation} \label{eq2}
\begin{split}
L_{PA} = & E_H[-y^p_H\log(G^{PA}_{O\LArrow{.15cm}H}(O)) + (1 -y^p_H)\log(1 - G^{PA}_{O\LArrow{.15cm}H}(O))] \\
& + E_O[-y^p_O\log(G^{PA}_{H\LArrow{.15cm}O}(H)) + (1 -y^p_O)\log(1 - G^{PA}_{H\LArrow{.15cm}O}(H))]
\end{split}
\end{equation}
$y^p_H$ and $y^p_O$ are the ground-truth labels for pathology samples. We aim to solve the following minmax optimization problem:
\begin{equation} \label{eq2}
\begin{split}
&G^{*}_{O\LArrow{.15cm}H}, G^{*}_{H\LArrow{.15cm}O}=
\\
& \arg \min\max L(G_{O\LArrow{.15cm}H}, G_{H\LArrow{.15cm}O}, D_H, D_O, G^{SC}_{O\LArrow{.15cm}H},  G^{SC}_{H\LArrow{.15cm}O}, G^{PA}_{O\LArrow{.15cm}H}, G^{PA}_{H\LArrow{.15cm}O})
\end{split}
\end{equation}

\section{Experiments}
\subsection{Experimental settings}
\subsubsection{Experimental dataset: }
Human coronary samples were collected from The University of Alabama at Birmingham. Specimens were imaged via a commercial OCT system (Thorlabs). A total of 194 OCT images were collected from 23 patients with an imaging depth of 2.56 mm. The pixel size was 2 µm × 2 µm within a B-scan. The width of the images ranged from 2 mm to 4 mm depending on the size of sample. After OCT imaging, samples were processed for H\&E histology at The University of Alabama at Birmingham. 
Among the dataset, 112 OCT images are from normal samples with the three-layer structure (i.e., intima, media, and adventitia); 82 OCT and H\&E images contain pathological patterns. 
The OCT and H\&E images are divided into non-overlap patches with a size of 368$\times$368. We randomly flip the patches from left to right for data augmentation. The training set contains 4297 OCT image patches and 4297 H\&E image patches.
\subsubsection{Implementation details: }
We adopt three convolution and transpose convolution layers with a stride of two for building a U-Net like structure for generating multi-scale feature maps. For the STB, we follow the design in \cite{9607618}. Our design of SCPA module is inspired by \cite{R.Strudel.2021}. But different from \cite{R.Strudel.2021}, we design the SCPA module to be capable of performing both segmentation and classification tasks, which suits our need for structural constraining and pathology awareness functions during virtual staining. The SCPAT-GAN is implemented by Pytorch. For training, the hyperparameters $\alpha$, $\beta$, $\gamma$, and $\iota$ are set to 1, 0.2, 5, and 5 empirically. The pixel values of OCT and H\&E images are scaled to [0, 1]. The batch size is 9. The learning rate is initialized as $10^{-4}$, followed by a linearly decaying decay for every 2 epochs. In total, the SCPAT-GAN is trained 10,000 epochs to ensure convergence. The experiments are carried out on an RTX A6000 GPU. 

\subsubsection{Metrics: }
We measure the similarity \textcolor{black}{of pairs of virtual histology and real histology images} using reference-free metrics including Fréchet inception distance (FID) \cite{Heusel.2017} and Perceptual hash value (PHV) \cite{Liu.2021}. We use the three variations of PHV scores ($i=1$, PHV1), ($i=2$, PHV2), and ($i=3$, PHV3) which are extracted from different levels $i$ of feature maps.
Also, we design a protocol to involve a pathologist to evaluate the quality of the virtual H\&E images. Real and virtual H\&E images are given to the pathologist, who is blinded to the true labels, to make predictions. 
We compare the prediction results from the pathologist with the true labels.

\subsection{Results and Discussion}

\subsubsection{Quantitative analysis: }
\begin{table}[t]
\centering
\caption{The FID and PHV scores of SCPAT-GAN, our previous method, and Cycle-GAN. The PHV scores calculated from different levels of the feature maps PHV$1$, PHV$2$, and PHV$3$. All the results are calculated using three-fold cross-validation.}
\label{table1}
\resizebox{0.7\textwidth}{!}{
\begin{tblr}{
  row{even} = {c},
  row{3} = {c},
  cell{1}{1} = {c},
  cell{1}{2} = {c=4}{c},
  cell{1}{6} = {c},
  cell{2}{2} = {c=4}{c},
  cell{2}{6} = {c},
  cell{2}{7} = {c},
  cell{2}{8} = {c},
  cell{3}{2} = {c=4}{},
  cell{4}{2} = {c=4}{},
  cell{5}{1} = {c},
  hline{1-2,5} = {-}{},
}
Metrics             & FID$\downarrow$ &  &  &  & PHV1$\uparrow$ & PHV2$\uparrow$ & PHV3$\uparrow$ \\
SCPAT-GAN           & \textbf{175.70}          &  &  &  & \textbf{57.41}          & \textbf{62.42}          & \textbf{52.93}          \\
GAN method in \cite{https://doi.org/10.48550/arxiv.2211.06737} & 238.74          &  &  &  & 48.18          & 55.17          & 49.29          \\
Cycle-GAN           & 284.53          &  &  &  & 45.48          & 54.31          & 48.67          \\
                    &                 &  &  &  &                &                &                
\end{tblr}}
\end{table}

The quantitative results of SCPAT-GAN for virtual H\&E are shown in Table \ref{table1}. Compared to our previous method and Cycle-GAN, the SCPAT-GAN generates virtual H\&E images of better quality, with lower FID scores and higher PHV scores. Those scores indicate that the virtual histology and real histology are perceptually similar. Moreover, we have a pathologist with more than 30 years of experience to evaluate the quality of virtual H\&E images. The pathologist, who is blind to the true labels, manually identifies if an image is real or virtual. 

\begin{figure}[t]
\centering
\includegraphics[width=12cm]{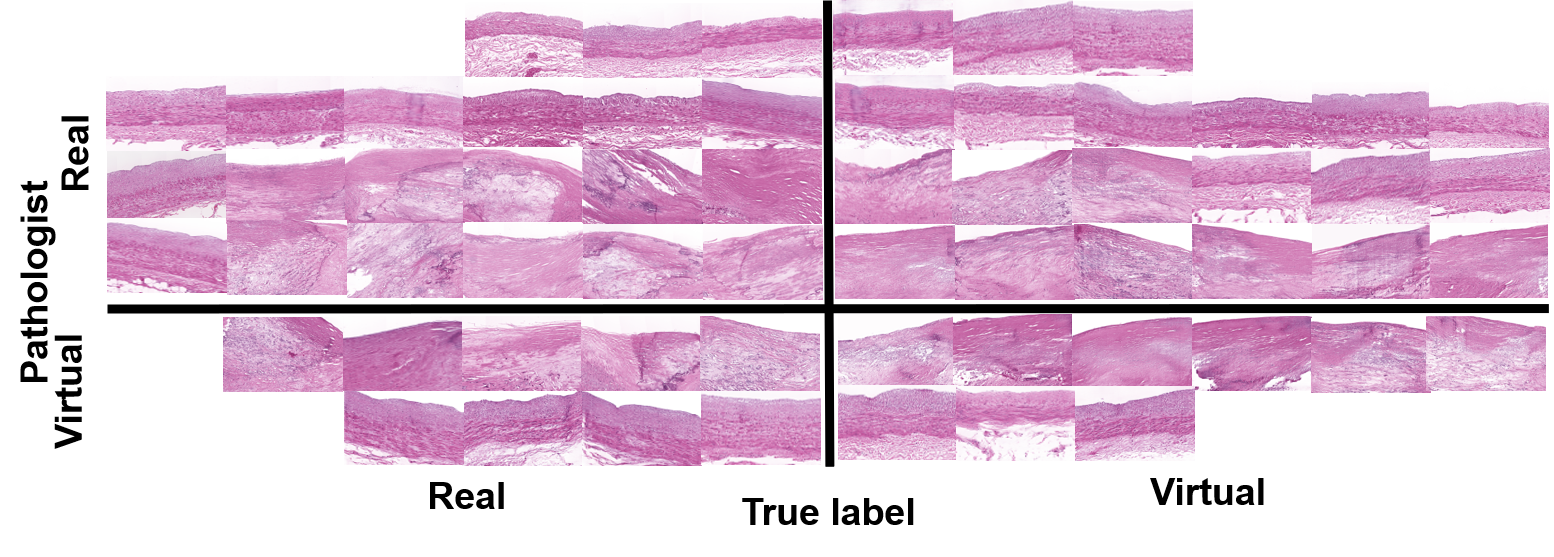}
\caption{Results of the pathologist's evaluation of real and virtual H\&E images. }
\label{fig:Conf}
\end{figure}

The results are shown in Figure \ref{fig:Conf}. 
\textcolor{black}{Among the total 60 images (half virtual and half real), over half of them (42 images) are deemed as 'real' by the pathologist. For the images that are deemed as 'virtual', half of them (9 images) are truly virtual while another half of images are real. This accuracy of the pathologist is close to the level of random guessing, which indicates that the generated virtual histology images are indistinguishable from real histology images.}

\begin{table}[t]
\centering
\caption{The ablation study. We disable pathological awareness module (SCT-GAN), structural constraining module (PAT-GAN), and both modules (T-GAN). }
\label{table2}
\resizebox{0.7\textwidth}{!}{
\begin{tblr}{
  row{even} = {c},
  row{3} = {c},
  row{5} = {c},
  cell{1}{1} = {c},
  cell{1}{2} = {c},
  cell{1}{3} = {c},
  cell{2}{2} = {c},
  cell{2}{3} = {c},
  cell{2}{4} = {c},
  cell{2}{5} = {c},
  hline{1-2,6} = {-}{},
}
Metrics   & FID$\downarrow$ & PHV1$\uparrow$ & PHV2$\uparrow$ & PHV3$\uparrow$ \\
SCPAT-GAN & \textbf{175.70}          & \textbf{57.41}          & \textbf{62.42}          & \textbf{52.93}          \\
SCT-GAN   & 177.60          & 53.59          & 59.28          & 51.33          \\
PAT-GAN   & 176.96          & 53.57          & 59.94          & 50.45          \\
T-GAN     & 203.24          & 47.00          & 54.32          & 47.48          
\end{tblr}}
\end{table}
\subsubsection{Ablation study: }
We perform an ablation study by disabling the structural constraining (PAT-GAN) or pathology awareness functions (SCT-GAN) or both (T-GAN) to justify the design of SCPAT-GAN as shown in Table \ref{table2}. When both structural constraining and pathological awareness functions are activated, SCPAT-GAN reaches the best performance.

\subsubsection{Qualitative analysis: }
\begin{figure}[t]
\centering
\includegraphics[width=12cm]{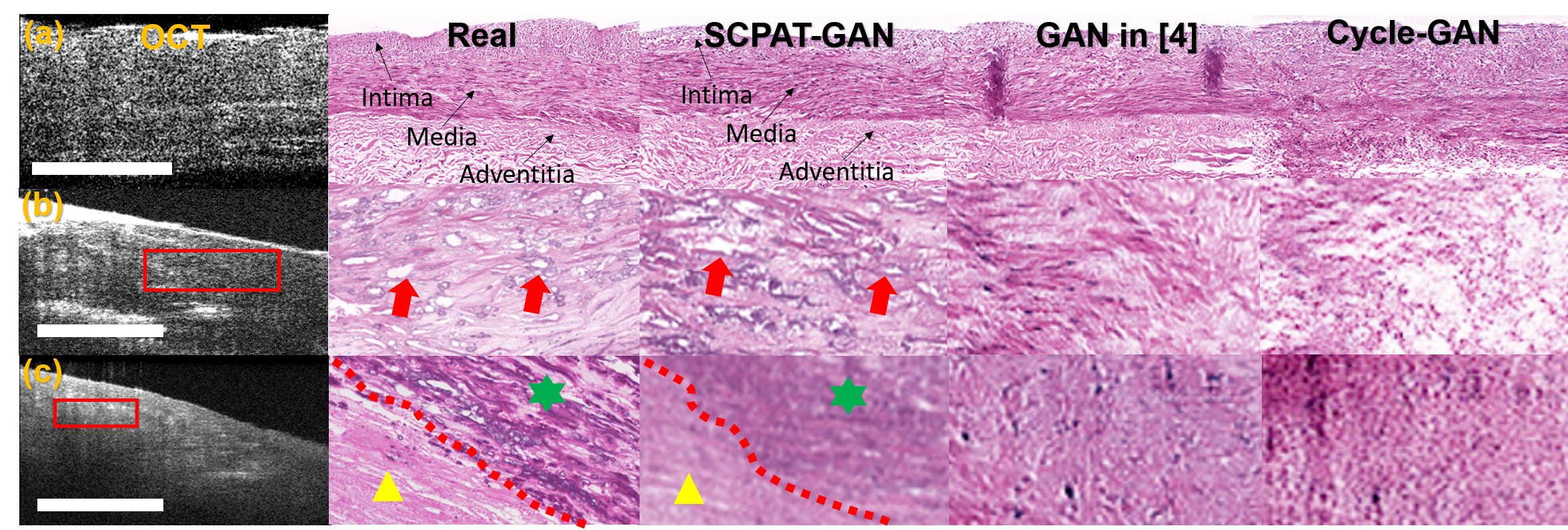}
\caption{Visual inspection of virtual H\&E images generated by SCPAT-GAN, GAN in \cite{https://doi.org/10.48550/arxiv.2211.06737}, and Cycle-GAN. (a): The normal coronary sample. The virtual H\&E image is very similar to the real H\&E image with the three-layer structure resolved. In virtual H\&E images, the lipid-rich regions appear as white holes. (b): Coronary samples with lipid-rich regions. (c): Coronary samples with calcified regions. The triangle and star represent different texture contrast. The calcified region has a color of dark purple. *The Contrasts of OCT images in (a)-(c) are enhanced to highlight the texture information for better visualization. The scale bar: 500 $\mu$m.}
\label{fig:VisualInspection}
\end{figure}

We visually inspect the virtual H\&E images generated by SCPAT-GAN in Figure \ref{fig:VisualInspection}. For normal coronary samples, the SCPAT-GAN is capable of generating the three-layer structure; for pathological coronary samples, the SCPAT-GAN is capable of resolving lipid-rich (red arrow) and calcified patterns (yellow star). Compared to real H\&E images, virtual H\&E images generated by SCPAT-GAN show similar patterns for lipid-rich and calcified regions. In the contrast, the GAN in \cite{https://doi.org/10.48550/arxiv.2211.06737} and Cycle-GAN fail to generate pathological patterns.

\section{Discussion and Conclusion}
In this paper, we design a novel convolutional transformer-GAN, namely SCPAT-GAN, for generating virtual H\&E histology from OCT images. Our SCPAT-GAN algorithm is capable of virtually staining OCT images for both normal and pathological human coronary samples. The SCPAT-GAN is weekly-supervised and does not require pixel-wisely matched OCT and H\&E datasets. By incorporating structural constraining and pathology awareness functions in the SCPAT-GAN, our methods outperform existing methods, which is confirmed by both objective metrics and the pathologist's evaluation. 
 
Compared to other label-free \cite{Cao.2023} or stain-to-stain \cite{Liu.2021} virtual histology works \cite{Bai.2023} which focused on top-view images, our SCPAT-GAN is designed for cross-sectional, depth-resolved OCT images and human coronary samples. Our SCPAT-GAN is the first to generate virtual H\&E images with pathological patterns for coronary samples based on OCT. In the future, we will further reducing the computational overhead of SCPAT-GAN via lightweight neural network \cite{belousov2021mobilestylegan}.

\subsubsection{Acknowledgements}
We would like to thank National Science Foundation
(CRII-1948540) and New Jersey Health Foundation.
\bibliographystyle{unsrt}
\bibliography{reference}

\begin{thebibliography}{10}

\bibitem{hajar2017risk}
Rachel Hajar.
\newblock Risk factors for coronary artery disease: historical perspectives.
\newblock {\em Heart views: the official journal of the Gulf Heart
  Association}, 18(3):109, 2017.

\bibitem{Tearney.2012}
Guillermo~J. Tearney, Evelyn Regar, Takashi Akasaka, et~al.
\newblock Consensus standards for acquisition, measurement, and reporting of
  intravascular optical coherence tomography studies: A report from the
  international working group for intravascular optical coherence tomography
  standardization and validation.
\newblock {\em Journal of the American College of Cardiology},
  59(12):1058--1072, 2012.

\bibitem{Winetraub.2021}
Yonatan Winetraub, Edwin Yuan, Itamar Terem, et~al.
\newblock {OCT2Hist}: Non-invasive virtual biopsy using optical coherence
  tomography.
\newblock {\em medRxiv}, 2021.

\bibitem{https://doi.org/10.48550/arxiv.2211.06737}
Xueshen Li, Hongshan Liu, Xiaoyu Song, Brigitta~C. Brott, Silvio~H. Litovsky,
  and Yu~Gan.
\newblock Structural constrained virtual histology staining for human coronary
  imaging using deep learning, 2022.

\bibitem{RFB15a}
O.~Ronneberger, P.Fischer, and T.~Brox.
\newblock {U-Net}: Convolutional networks for biomedical image segmentation.
\newblock In {\em Medical Image Computing and Computer-Assisted Intervention
  (MICCAI)}, volume 9351 of {\em LNCS}, pages 234--241. Springer, 2015.
\newblock (available on arXiv:1505.04597 [cs.CV]).

\bibitem{9607618}
Jingyun Liang, Jiezhang Cao, Guolei Sun, Kai Zhang, Luc Van~Gool, and Radu
  Timofte.
\newblock {SwinIR}: Image restoration using swin transformer.
\newblock In {\em 2021 IEEE/CVF International Conference on Computer Vision
  Workshops (ICCVW)}, pages 1833--1844, 2021.

\bibitem{8237506}
Jun-Yan Zhu et~al.
\newblock Unpaired image-to-image translation using cycle-consistent
  adversarial networks.
\newblock In {\em 2017 IEEE International Conference on Computer Vision
  (ICCV)}, pages 2242--2251, 2017.

\bibitem{Liu.2021}
Shuting Liu, Baochang Zhang, Yiqing Liu, Anjia Han, Huijuan Shi, Tian Guan, and
  Yonghong He.
\newblock Unpaired stain transfer using pathology-consistent constrained
  generative adversarial networks.
\newblock {\em IEEE transactions on medical imaging}, 40(8):1977--1989, 2021.

\bibitem{R.Strudel.2021}
{R. Strudel}, {R. Garcia}, {I. Laptev}, and {C. Schmid}.
\newblock Segmenter: Transformer for semantic segmentation.
\newblock In {\em 2021 IEEE/CVF International Conference on Computer Vision
  (ICCV)}, pages 7242--7252, Los Alamitos, CA, USA, 2021. {IEEE Computer
  Society}.

\bibitem{Heusel.2017}
Martin Heusel, Hubert Ramsauer, Thomas Unterthiner, Bernhard Nessler, and Sepp
  Hochreiter.
\newblock {GANs} trained by a two time-scale update rule converge to a local
  nash equilibrium.
\newblock In {\em Proceedings of the 31st International Conference on Neural
  Information Processing Systems}, NIPS'17, pages 6629--6640, Red Hook, NY,
  USA, 2017. {Curran Associates Inc}.

\bibitem{Cao.2023}
Rui Cao, Scott~D. Nelson, Samuel Davis, Yu~Liang, Yilin Luo, Yide Zhang, Brooke
  Crawford, and Lihong~V. Wang.
\newblock Label-free intraoperative histology of bone tissue via
  deep-learning-assisted ultraviolet photoacoustic microscopy.
\newblock {\em Nature Biomedical Engineering}, 7(2):124--134, 2023.

\bibitem{Bai.2023}
Bijie Bai, Xilin Yang, Yuzhu Li, Yijie Zhang, Nir Pillar, and Aydogan Ozcan.
\newblock Deep learning-enabled virtual histological staining of biological
  samples.
\newblock {\em Light: Science {\&} Applications}, 12(1):57, 2023.

\bibitem{belousov2021mobilestylegan}
Sergei Belousov.
\newblock Mobilestyle{GAN}: A lightweight convolutional neural network for
  high-fidelity image synthesis.
\newblock {\em arXiv preprint arXiv:2104.04767}, 2021.

\end{thebibliography}
\end{document}